\documentclass{article}
\usepackage{amsmath,amssymb}

\newcommand{\bs}{\boldsymbol}
\title{Classical Dynamics in Deformed Spaces}
\author{A.\ Leznov and J.\ Mostovoy}
\date{}

\begin{document}

\maketitle

\begin{abstract}
We consider a 3-parametric linear deformation of the Poisson
brackets in classical mechanics. This deformation can be thought of as
the classical limit of dynamics in so-called "quantized spaces".
Our main result is a description of the motion of a particle in the
corresponding Kepler-Coulomb problem.
\end{abstract}

\section{Introduction}

The set-up of a problem in quantum mechanics usually consists of
specifying (a) the structure of the algebra of the observables and
(b) the Hamiltonian of the problem. If the commutation relations in
the algebra of the observables are deformations of the usual
commutation relations between the coordinates and the momenta,
the situation is often referred to as "dynamics in a quantized space".
Schr\"odinger was apparently the first to consider such a situation:
in \cite{Schr} he calculated the energy levels for the Coulomb potential
on a 3-sphere. For this problem the term "quantized space" is rather
appropriate as the spectrum of the distance operator on a 3-sphere
is discrete even in the absence of any field.

One can obtain "classical dynamics in a quantized space" by setting
\[\lim \frac{i}{h} [A,B]=\{A,B\}\quad \text{as}\quad  h\to 0,\]
where $\{\cdot,\cdot\}$ are the Poisson brackets of classical dynamics.
We prefer to speak about "classical dynamics in deformed spaces"
(keeping in mind that it is not just the space but the whole dynamics
being deformed).

In fact, one does not need a reference to the quantum theory in order to
construct a deformation of the functional algebra of classical
observables. However, we are specifically interested in the classical
version of the 3-parametric family of deformations considered in
\cite{LK, Leznov, KL} in the context of quantum mechanics.

This family arises naturally if one considers deformations of
the commutation relations involving the four space-time coordinates, ten
generators of the Poincare algebra (that is, infinitesimal translations and
proper Lorentz transformations) and the unity element, with the property
that the resulting algebra contains the Lorentz algebra as a subalgebra.

In what follows we will consider only three-dimensional non-relativistic
problems. The usual commutation relations between the dynamic variables
are replaced by the following set of brackets satisfying the Jacobi identity:
\begin{equation}
\label{commrel}
\begin{array}{lll}
\{p_i,x_j\}=\delta_{ij}I-\frac{\epsilon_{ijk}l_{k}}{S},
&\{l_i,x_j\}=-\epsilon_{ijk}x_{k},
&\{I,x_i\}=\frac{x_i}{S}-\frac{p_i}{M^2},  \\
\{x_i,x_j\}=-\frac{\epsilon_{ijk}l_{k}}{M^2},
&\{l_i,p_j\}=-\epsilon_{ijk}p_{k},
&\{I,p_i\}=\frac{x_i}{L^2}-\frac{p_i}{S},\\
\{p_i,p_j\}=-\frac{\epsilon_{ijk}l_{k}}{L^2},
&\{l_i,l_j\}=-\epsilon_{ijk}l_{k},
&\{I,l_i\}=0. \\
\end{array}
\end{equation}
Here $x_i$ are the coordinates and $p_i$ and $l_i$ are the components of
the momentum and angular momentum respectively; the function $I$ is a
deformation of the unity function. The three parameters $L^2$, $M^2$ and
$S$ have the dimensions of area, square of the momentum and action
respectively.

In general, these parameters may depend on rotation-invariant functions
of the dynamic variables such as $x^2,l^2,\bs{(px)}$ etc.
In this case in the resulting non-linear functional algebra
the Jacobi identities are still satisfied. Here, however, we only
consider the case of $L^2,M^2$ and $S$ being constants. It should be
stressed that we do not require the signs of $L^2$ and $M^2$
to be positive.

In order to obtain the flat space with the usual dynamics on the scale of
the Solar system one has to assume that the values of $L^2$, $M^2$ and $S$
are huge. In fact, taking $M^2$ and $S$ to be infinite, we get the
dynamics in the space of constant curvature equal to $1/L^2$, so $L$ is
on the scale of the size of the Universe.

An important feature of the relations (\ref{commrel}) is their non-invariance
with respect to the time reversal. Namely, changing the signs of
$\bs{p}$, $\bs{l}$ and of the brackets simultaneously one obtains
relations of the same form but with the sign of the parameter $S$ reversed.

Various particular cases of the above commutation relations are studied in the
literature. The case of the space of constant curvature (that is,
$M^2,S=\infty$) has received most attention. The first work related to this
case is due to Schr\"odinger \cite{Schr} (who assumed the curvature to be
positive, for the case of negative curvature see \cite{IS}). Many
other works on the subject have appeared since; we only mention \cite{Higgs}
which contains a discussion of the classical mechanics (and, in particular,
the Kepler problem) in the space of constant curvature.

The case $S=\infty$ is due to Yang \cite{Yang}. 
The general case of the above commutation relations was
introduced in \cite{LK}, see also \cite{KL}. The quantum energy levels for
the Kepler-Coulomb potential and the harmonic oscillator for this general
case are calculated in \cite{Leznov}.

Other approaches to deformed spaces exist: see, for example, \cite{CZ}.

About the notation: we use bold italic letters for vectors. The scalar
and vector products of vectors $\bs{a}$ and $\bs{b}$ will be denoted by
$\bs{(ab)}$ and $\bs{[ab]}$ respectively, the absolute value of $\bs{a}$
is denoted by $|\bs{a}|$ and its square - by $a^2$.

For simplicity, we always assume the particles to be of unit mass.

\section{The canonical coordinates}

\subsection{The canonical coordinates}

Depending on the signs of the constants $L^2$, $M^2$ and $L^2M^2-S^2$,
the functional algebra (\ref{commrel}) is equivalent as a Lie algebra
to one of the algebras $o(5)$, $o(1,4)$ or $o(2,3)$ (\cite{Leznov}).
According to the Darboux theorem these algebras can be resolved in terms
of four pairs of independent canonically conjugate coordinates and momenta
and two cyclic variables constructed from the elements of
(\ref{commrel}). In the algebraic language, the cyclic variables are
the Casimir operators of the corresponding algebra.

The Casimir operators of the algebra under consideration
are as follows:
\[K_2=I^2+\frac{x^2}{L^2}+\frac{p^2}{M^2}-\frac{2\bs{(px)}}{S}-
l^2\left(\frac{1}{S^2}-\frac{1}{L^2M^2}\right)\]
and
\[K_4= (I\bs{l}-\bs{[x,p]})^2-\frac{\bs{(pl)}^2}{M^2}+
\frac{2\bs{(pl)(xl)}}{S}-\frac{\bs{(xl)}^2}{L^2}.\]
The Poisson brackets of $K_2$ and $K_4$ with any function vanish
and their values can be fixed. In order to get a correct limit in the usual
(that is, nondeformed) case we shall set $K_2=1$. As for the value
of $K_4$, we will only consider the scalar representation of the
algebra of the observables, that is, set
\begin{equation}
\label{l}
I\bs{l}=\bs{[xp]}.
\end{equation}
This implies that $\bs{(pl)}=\bs{(xl)}=0$ and, hence, that $K_4=0$.
The relation (\ref{l}) generalises the definition of the angular
momentum as a function of other variables to the case of the deformed
space.

Imposing the relation (\ref{l}) we reduce the number of Darboux pairs
by one. An explicit resolution of the algebra (\ref{commrel}) in terms of
three angular variables can be found in the Appendix.

The equality $K_2=1$ (with $l^2$ expressed via $\bs{p}$, $\bs{x}$ and $I$)
can be thought of as the equation defining the phase space of our dynamical
system. In particular, if $M^2,S=\infty$ and $L^2>0$ this is the equation of
$S^3\times{\bf R}^3$ in ${\bf R}^7$ and the relations $(\ref{commrel})$
describe the dynamics on the 3-sphere of radius $L$. In fact, the phase
space is a manifold whenever $S^2=L^2M^2$. In general, however, it is not
smooth: if $S^2\neq L^2M^2$ there is a singularity at $I=0$.

\subsection{Problems with spherical symmetry}\label{sss}

For problems with spherical symmetry (such as the Kepler problem considered
below) it may be of use considering the algebra of observables invariant
under rotations. Namely, consider the algebra generated by the 5 spherically
invariant variables:
$r^2\equiv x^2+\frac{l^2}{M^2}$, ${p'}^2\equiv p^2+
\frac{l^2}{L^2}$, $D\equiv \bs{(px)}+\frac{l^2}{S}$, $I$,
 $l^2$. The commutation relations of the nonlinear algebra
generated by these 5 variables follow directly from (\ref{commrel}):

\begin{equation}
\begin{array}{lll}\label{IA}
\{I,{p'}^2\}=2\left(\frac{D}{L^2}-\frac{{p'}^2}{S}\right),&
\{I,{r}^2\}=-\left(\frac{D}{M^2}-\frac{r^2}{S}\right),&
\{I,D\}=\frac{r^2}{L^2}-\frac{{p'}^2}{M^2}\\
\{{p'}^2,r^2\}=4DI,&
\{D,{p'}^2\}=-2{p'}^2I,&
\{D,r^2\}=2r^2I
\end{array}
\end{equation}

The nonlinear functional algebra (\ref{IA}) is generated by 5 elements and
has three cyclic variables: two of them, namely $K_2$ and $K_4$, come from
the initial algebra. 
The third cyclic variable is $l^2$ which commutes with all the generators
of (\ref{IA}). Thus, by the Darboux theorem, this algebra can be resolved
in terms of one coordinate and one momentum. We choose $r$ as the
coordinate. (In the non-deformed case this is just the radial variable.)

The values of $K_2$ and $K_4$ being fixed, the quantities $D$, $p'$ and $I$
depend only on $r$, the corresponding canonically conjugate momentum
which we denote by $\rho$, and $l^2$.
Taking the squares of the vectors on both sides of (\ref{l}) and using the
fact that $K_2=1$ after some manipulations we obtain
\begin{equation}\label{pshtrih}
{p'}^2=\frac{1}{r^2}\left(D^2+l^2\right).
\end{equation}
From the last equality of (\ref{IA}) we get that
\[ I=\frac{1}{r}D_{\rho} \]
Substituting this expression together with (\ref{pshtrih}) into the equality
$K_2=1$ we obtain an equation on $D$
\[
1=\frac{D_{\rho}^2}{r^2}+\frac{D^2+l^2}{M^2r^2}+\frac{r^2}{L^2}-
\frac{2D}{S}
+l^2\left(\frac{1}{S^2}-\frac{1}{L^2M^2}\right)
\]
with the general solution
\begin{equation}\label{d}
D=\frac{M^2r^2}{S}-\sqrt {\left(1+r^2M^2\left(\frac{1}{S^2}-
\frac{1}{L^2M^2}\right)\right) (M^2r^2-l^2)}\cos{\frac{\rho}{M}}.
\end{equation}
As a corollary we have
\begin{equation}\label{i}
I= \sqrt {\left( 1+r^2M^2\left(\frac{1}{S^2}-
\frac{1}{L^2M^2}\right)\right) \left(1-\frac{l^2}{M^2r^2}\right)}\sin{\frac{\rho}{M}}
\end{equation}
Thus all elements of the nonlinear functional algebra given by (\ref{IA})
are expressed in terms of the canonically conjugate pair $(r,\rho)$ and
the square of the angular momentum $l^2$.

\section{Free motion}

The Hamiltonian of the free motion is chosen so as to commute with
infinitesimal translations and rotations, that is, with $\bs{p}$ and $\bs{l}$.
The simplest such expression which gives the correct limit as the parameters
tend to infinity is
\[E=\frac{1}{2}\left(p^2+\frac{l^2}{L^2}\right) .\]
It follows immediately from the commutation relations that the vectors
$\boldsymbol{p}$ and $\boldsymbol{l}$ are constant. Also,
\[\stackrel{\cdot}{\boldsymbol{x}}=I\boldsymbol{p}
-\frac{\boldsymbol{[xl]}}{L^2}+
\frac{\boldsymbol{[pl]}}{S}\]
and
\[\stackrel{\cdot}{I}=\frac{p^2}{S}-\frac{\boldsymbol{(px)}}{L^2}.\]
From this
\[\stackrel{\cdot\cdot}{I}+\frac{2E}{L^2}I=0\]
and, hence, \[I=I_0\cos{\left(\frac{\sqrt{2E}}{L}(t-t_0)\right)}.\]
Now, taking the vector product of (\ref{l}) with $\boldsymbol{p}$
one has
\[\boldsymbol{(px)p}-p^2\boldsymbol{x}=I\boldsymbol{[lp]}\]
from where
\begin{equation}
\label{fm}
\boldsymbol{x}=\left(\frac{L^2}{S}+\frac{I_0}{p^2}L\sqrt{2E}
\sin{\left(\frac{\sqrt{2E}}{L}(t-t_0)\right)}\right)\boldsymbol{p}-
\frac{I_0}{p^2}\cos{\left(\frac{\sqrt{2E}}{L}(t-t_0)\right)}\boldsymbol{[lp]}.
\end{equation}
The value of $I_0$ can be found from the condition $K_2=1$:
\[I_0=\frac{p}{\sqrt{2E}}\cdot
\sqrt{1+2E\left(\frac{L^2}{S^2}-\frac{1}{M^2}\right)}.\]
In the case $L^2=\infty$ the expression for $\bs{x}$
should be modified to read
\[\bs{x}=\bs{x}_0+\left(I_0\bs{p}+\frac{\bs{[pl]}}{S}\right)t+\frac{p^2}{2S}\bs{p}t^2\]
with
\[I_0=\sqrt{1-\frac{p^2}{M^2}+\frac{l^2}{S^2}+\frac{2\bs{(px_0)}}{S}}.\]

The orbits of the free motion are elliptic, parabolic or hyperbolic
according to whether $L^2$ is positive, zero, or negative. This should not
be surprising given that $L^{-2}$ can be interpreted as the curvature of the
space. For example, set $M^2,S=\infty$ and take $L^2>0$. Then the
elliptic orbits (\ref{fm}) can be obtained from the geodesics
on the 3-sphere
\[1=I^2+\frac{x^2}{L^2}\]
by projecting them orthogonally on a Euclidean 3-space. In particular,
the farthest distance from the origin reached by a freely moving particle
is the same for all orbits and equals to $L$.


\section{The Kepler problem}
\subsection{The equation of the orbit}

The Hamiltonian for the Kepler problem is chosen (as in \cite{Leznov})
so as to commute with the R\"unge-Lenz vector $\bs{A}$ defined as
\begin{equation}
\label{rl}
\bs{A}=\bs{[pl]}+\frac{\alpha\bs{x}}{r}=
\frac{1}{I}\left(p^2\bs{x}-\bs{(px)p}\right)+\frac{\alpha\bs{x}}{r}
\end{equation}
where $r$ is given by $r^2=x^2+\frac{l^2}{M^2}$.

The commutation relations for the components of $\bs{A}$ are as follows:
\[\{A_i,A_j\}=2(E-\frac{l^2}{L^2})\epsilon_{ijk}l_{k},\]
where $E$ stands for the expression
\begin{equation}\label{H}
E=\frac{1}{2}\left(p^2+\frac{l^2}{L^2}+\frac{2\alpha I}{r}-
\frac{\alpha^2}{M^2 r^2}\right).
\end{equation}
Also,
\[\{l_i,A_j\}=-\epsilon_{ijk}A_{k}\]
and, hence,
\[\{l^2,A_j\}=2\epsilon_{ijk}l_{j}A_{k}.\]
To see that $E$ commutes with all of the components of $\bs{A}$
first notice that
\begin{equation}\label{E}
A^2=\alpha^2 + l^2 (2E-\frac{l^2}{L^2}).
\end{equation}
Now,
\[2\{E,A_i\}=\{\frac{A^2}{l^2},A_i\}-\alpha^2\{\frac{1}{l^2},A_i\}+
\frac{1}{L^2}\{l^2,A_i\}=0.\]
Thus $E$ can be chosen to be the Hamiltonian for the Kepler problem.

The three vectors $\bs{x}$, $\bs{p}$ and $\bs{A}$ lie in the plane orthogonal
to the vector of angular momentum. Let us choose the polar coordinates
$(|\bs{x}|,\phi)$ in this plane with the direction $\phi=0$
being given by the R\"unge-Lenz vector. (If $\bs{A}=0$ we take $\phi=0$
to be an arbitrary ray passing through the origin.)

Taking the scalar product of (\ref{rl}) with $\bs{x}$ one has
\begin{equation}
\label{I}
\frac{A|\bs{x}|\cos{\phi}-\alpha r}{l^2} = I- \frac{\alpha}{M^2 r}
\end{equation}
Taking the vector product of $\bs{A}$ and $\bs{l}$ we have
\begin{equation}
\label{xl}
\bs{[Al]}=-l^2\bs{p}+\frac{\alpha}{r}\bs{[xl]}.
\end{equation}
The scalar product of (\ref{xl}) with $\bs{x}$ gives
\begin{equation}
\label{xldp}
\bs{(px)}=-\frac{1}{l^2}(\bs{[Al]},\bs{x})=
\frac{A|\bs{x}|\sin{\phi}}{l}.
\end{equation}
Now the orbits can be found as follows. In the expression $K_2=1$
one can substitute $\bs{(px)}$ from (\ref{xldp}), $I$ --- from
(\ref{I}) and $p^2$ --- from (\ref{H}) and (\ref{E}). The resulting
equation of the orbit is as follows:
\begin{equation}
\label{orbit}
1+\frac{2A|\bs{x}|\sin{\phi}}{lS}=\frac{(A|\bs{x}|\cos{\phi}-\alpha r)^2}{l^4}
+\frac{r^2}{L^2}+\frac{A^2-\alpha^2}{l^2 M^2}-
\frac{l^2}{S^2}.
\end{equation}

In the Cartesian coordinates
$(x,y)=(|\bs{x}|\cos{\phi}, |\bs{x}|\sin{\phi})$
the equation (\ref{orbit}) takes the following form:

\begin{multline} \label{quartic}
x^2\left(\frac{A^2+\alpha^2}{l^4}+\frac{1}{L^2}\right)
+y^2\left(\frac{\alpha^2}{l^4}+\frac{1}{L^2}\right)-\frac{2Ay}{lS}+
\frac{A^2}{l^2 M^2}-
l^2\left(\frac{1}{S^2}-\frac{1}{L^2M^2}\right)-1\\
=\frac{2\alpha A}{l^4}x\sqrt{x^2+y^2+\frac{l^2}{M^2}}.
\end{multline}

This equation defines a branch of a plane quartic. The other branch
of this quartic describes the orbit with the same values of
$A$ and $\bs{l}$ but with the opposite value of $\alpha$.
Notice that changing $\bs{l}$ for $\bs{-l}$ we obtain, generally, a
different orbit. (Recall that the commutation relations (\ref{commrel})
are not invariant under time reversal.)

Interestingly, it was Giovanni Domenico Cassini (1625--1712) who first
suggested that the trajectories of planets are quartic curves.  Now these
curves are known as Cassinian ovals. A Cassinian oval is defined
as the locus of points with the product of their distances
from two fixed points equal to a fixed constant.


\subsection{The parametric form for the orbit}

From now on we will assume for simplicity that $L^2$ is positive.
Taking into the account the fact that $K_2=1$, this implies, in particular,
that all orbits are closed.

The curve (\ref{quartic}) can be parametrised as follows.
The condition $L^2>0$ means that we can choose $a,b$ so that
\[a^2+b^2=\frac{\alpha^2+A^2}{l^4}+\frac{1}{L^2} \]
and
\[ab=\frac{A\alpha}{l^4}.\]
Explicitly,
\[a^2,b^2=\frac{1}{2}\left(\frac{\alpha^2+A^2}{l^4}+\frac{1}{L^2}\pm
\sqrt{\left(\frac{\alpha^2+A^2}{l^4}+\frac{1}{L^2}\right)^2-
\frac{4A^2\alpha^2}{l^8}}
\right)\]
and we take $a^2>b^2$.

The equation (\ref{quartic}) can be written as
\[(ax-br)^2+\left(\nu y-\frac{A}{lS\nu}\right)^2=\gamma^2.\]
Here
\[\nu^2
=\frac{1}{2}\left(\frac{\alpha^2-A^2}{l^4}+\frac{1}{L^2}+
\sqrt{{\left(\frac{\alpha^2+A^2}{l^4}+\frac{1}{L^2}\right)^2-
\frac{4A^2\alpha^2}{l^4}}}\right)\]
and
\[\gamma^2=1+l^2\left(\frac{1}{S^2}-\frac{1}{L^2M^2}\right)
\frac{a^2}{\nu^2}.\]
In the non-deformed limit $\gamma=1$ and
\[\nu^2=\frac{\alpha^2-A^2}{l^4}=-\frac{2E}{l^2}\]
for $E<0$ and $\nu^2=0$ otherwise.

We can set
\begin{align*}
ax-br&=\gamma\cos{\theta},\\
\nu y-\frac{A}{lS\nu}&=\gamma\sin{\theta}.
\end{align*}
so that
\begin{equation}\label{xy}
\begin{split}
r&=\frac{1}{a^2-b^2}\left(b\gamma\cos{\theta}
+|a|\sqrt{\gamma^2\cos^2{\theta}+(a^2-b^2)
\left(\left(\frac{\gamma}{\nu}\sin{\theta}+\frac{A}{lS\nu^2}\right)^2+
\frac{l^2}{M^2}\right)}\right),\\
x&=\frac{1}{a^2-b^2}\left(a\gamma\cos{\theta}
+b\frac{|a|}{a}\sqrt{\gamma^2\cos^2{\theta}+(a^2-b^2)
\left(\left(\frac{\gamma}{\nu}\sin{\theta}+\frac{A}{lS\nu^2}\right)^2+
\frac{l^2}{M^2}\right)}\right),\\
y&=\frac{\gamma}{\nu}\sin{\theta}+\frac{A}{lS\nu^2}.
\end{split}
\end{equation}
In the classical limit the above parametrisation
gives the usual trigonometric parametrisation of the elliptic
trajectory. There exists also a generalisation of the hyperbolic
parametrisation.  Write (\ref{quartic}) as
\[(bx-ar)^2-\left(\mu y+\frac{A}{lS\mu}\right)^2=\delta^2\]
with
\begin{equation}\label{mu}
\mu^2=-\frac{1}{2}\left(\frac{\alpha^2-A^2}{l^4}+\frac{1}{L^2}-
\sqrt{{\left(\frac{\alpha^2+A^2}{l^4}+\frac{1}{L^2}\right)^2-
\frac{4A^2\alpha^2}{l^4}}}\right)
\end{equation}
and
\[ \delta^2=1-l^2\left(\frac{1}{S^2}-\frac{1}{L^2M^2}\right)
\frac{b^2}{\mu^2}.\]
Assume that $\delta^2>0$. Then setting
\begin{align*}
bx-ar&=\delta\cosh{\theta},\\
\mu y+\frac{A}{lS\mu}&=\delta\sinh{\theta}.
\end{align*}
we obtain the hyperbolic parametrisation of the orbit.
(The case $\delta^2<0$ is entirely similar.)

Notice that $\mu^2\nu^2=\frac{A^2}{l^4L^2}$ so in the non-deformed limit
only one of the parametrisations is meaningful for the given values
of $\alpha$ and $A$, the other just giving the equation of the orbit.
In the deformed case both parametrisations can be meaningful
for the same closed orbit. In this case the hyperbolic parametrisation
is two-valued  and the parameter $\theta$ belongs to a closed interval.

In the non-deformed limit the hyperbolic parametrisation remains two-valued
with the parameter $\theta$ varying over all real numbers. The two
branches correspond to the two possible values of $I=\pm 1$.
One branch gives the usual parametrisation of a hyperbolic trajectory.
The second
branch parametrises a trajectory with the same value of $A$ and $l$
but with the opposite sign of $\alpha$.

\subsection{Dependence on time}

Taking the Poisson bracket of $\bs{x}$ with the
Hamiltonian we obtain the equations of
motion:
\[\stackrel{\cdot}{\boldsymbol{x}}=I\boldsymbol{p}
-\frac{\boldsymbol{[xl]}}{L^2}+
\frac{\boldsymbol{[pl]}}{S}+
\frac{\alpha\bs{x}}{Sr}-
\frac{\alpha\bs{p}}{M^2r}.\]
Substituting $I$ and $\bs{p}$ from (\ref{I}) and (\ref{xl}) respectively, we get
\[\stackrel{\cdot}{\boldsymbol{x}}=\frac{1}{l^4}(Ax-\alpha r)
\left(\frac{\alpha}{r}\bs{[xl]}-\bs{[Al]}\right)
-\frac{\boldsymbol{[xl]}}{L^2}+
\frac{\boldsymbol{A}}{S}.\]
In particular, for the coordinate $y$ we have:
\[\stackrel{\cdot}{y}=\frac{1}{l^3}\left(\left(A^2+\alpha^2+\frac{l^4}{L^2}\right)x-
\alpha A\left(\frac{x^2}{r}+r\right)\right),\]
or, equivalently,
\[\stackrel{\cdot}{y}=-\frac{l}{r}(ax-br)(bx-ar).\]
It follows from (\ref{xy}) that
\[\frac{r}{bx-ar}=-\frac{|a|}{a}\cdot\left(\frac{b\gamma\cos{\theta}}{\sqrt{\Delta}}+|a|\right)
\cdot\frac{1}{a^2-b^2}\]
where $\Delta$ is the expression under the square root in (\ref{xy}).
Thus
\begin{equation}\label{eq}
\frac{|a|}{a}\cdot l\nu (a^2-b^2)\cdot dt=
d\theta\left(|a|+\frac{b\gamma\cos{\theta}}{\sqrt{\Delta}}\right).
\end{equation}
The function $\Delta$ only depends on $\sin{\theta}$ so the period
of the motion can be found without integrating (\ref{eq}):
\[T=\frac{2\pi |a|}{l\nu (a^2-b^2)}.\]
Notice that for $A,l$ and $\alpha$ fixed, the period only depends on $L^2$
and not on $M^2$ and $S$.

One can re-write (\ref{eq}) as
\[
\frac{|a|}{a}\cdot(a^2-b^2)l dt=
\frac{|a|}{\nu}d\theta+\frac{b}{\mu}\frac{dq}{\sqrt{q^2+1-l^2\left(
\frac{1}{S^2}-\frac{1}{L^2M^2}\right)(\nu^2L^2-1)}}
\]
where
\[ q=\frac{\mu}{\nu}\gamma\sin{\theta}+\frac{(a^2-b^2)lL}{\nu S}\]
and $\mu$ is defined as in (\ref{mu}) so that $\mu^2+\nu^2=a^2-b^2$.
Integrating, one gets the following parametrisation of the time $t$ by
$\theta$:
\begin{equation}\label{answer}
\frac{|a|}{a}\cdot(t-t_0)(a^2-b^2)l =
\frac{|a|}{\nu}\theta+\frac{b}{\mu}
\sinh^{-1}
{
\left(
\frac
{
\frac{\mu}{\nu}\gamma\sin{\theta}+\frac{(a^2-b^2)lL}{\nu S}
}
{\sqrt{
1+l^2\left(\frac{1}{S^2}-\frac{1}{L^2M^2}\right)
(1-\nu^2 L^2)}}
\right).
}
\end{equation}
(The above answer is valid if
\[1+l^2\left(\frac{1}{S^2}-\frac{1}{L^2M^2}\right)(1-\nu^2 L^2)>0;\]
obviuos modifications are needed otherwise in (\ref{answer}).)

The standard trigonometric parametrisation of $t$ by $\theta$ in the
non-deformed Kepler problem can be obtained from (\ref{answer}) by setting
first $S$ and $M^2$ to be infinite and then taking the limit as
$L^2$ tends to infinity. (Note that as $L^2$ tends to $\infty$, the
parameter $\mu$ tends to 0, as $\mu=\frac{A}{\nu lL}$.)

\section{The Hamilton-Jacobi equation}
The spherically symmetric part of the Hamilton-Jacobi equation is obtained
by substituting (\ref{pshtrih}), (\ref{d}) and (\ref{i}) into the explicit
expression for the Hamiltonian function
\[
2E=\frac{1}{r^2}(D^2+l^2)+\frac{2\alpha}{r}I-\frac{\alpha^2}{M^2r^2}.
\]
Namely, we obtain
\begin{equation*}
\begin{split}
2Er^2-l^2+\frac{\alpha^2}{M^2}=\left(\frac{M^2r^2}{S}-
\sqrt{\left(1+r^2M^2\left(\frac{1}{S^2}-\frac{1}{L^2M^2}\right)\right)
(M^2r^2-l^2)}\cos{\frac{W_r}{M}}\right)^2
\\
+\frac{2\alpha}{M} \sqrt{\left(1+r^2M^2\left(\frac{1}{S^2}-\frac{1}{L^2M^2}\right)
\right) (M^2r^2-l^2)}\sin{\frac{W_r}{M}}
\end{split}
\end{equation*}
where $W$ is the radial part of the action function. One arrives to the
same expression considering the quasi-classical limit for the Schr\"odinger
equation in a deformed (quantized) space.

We do not know a direct method of solving this (rather unusual at the first
sight) equation. In fact, one only needs to know the the second derivative
$W_{rE}$ in order to describe $r$ as a function of time:
$r(t)$ can be found from the equality
\[ t-t_0=\int dr W_{rE}.\]
Below we obtain an algebraic equation for the second derivative
$W_{rE}$ of the radial part of the action.

Let us introduce the following notation:
\[
u=\frac{M^2r^2}{S}-\sqrt{\left(1+r^2M^2\left(\frac{1}{S^2}-\frac{1}{L^2M^2}
\right)\right)
(M^2r^2-l^2)}\cos{\frac{W_r}{M}}
\]
\[
v=\sqrt {\left(1+r^2M^2\left(\frac{1}{S^2}-
\frac{1}{L^2M^2}\right)\right)(M^2r^2-l^2)}
\sin{\frac{W_r}{M}}-\frac{\alpha}{M}
\]
In this notation the Hamilton-Jacobi equation becomes
\begin{equation}
u^2+\frac{2\alpha}{M}v=2Er^2-l^2-\frac{\alpha^2}{M^2}\equiv \theta. \label{Q1}
\end{equation}
Taking the derivative of both sides with respect to $E$ we have
\begin{equation}
uv=Mr^2\left(\frac{1}{W_{rE}}-\frac{\alpha}{S}\right)\equiv F\label{Q2}
\end{equation}
Finally, taking into the account that $\sin^2{x}+\cos^2{x}=1$ we obtain
\begin{equation}
v^2-\frac{2M^2r^2}{S}u=M^2r^2\left(1-\frac{2E}{M^2}-\frac{r^2}{L^2}-l^2\left(
\frac{1}{S^2}-\frac{1}{M^2L^2}\right)\right)\equiv \bar \theta\label{Q3}
\end{equation}
The condition of self-consistensy of the three equations above with respect
to only two functions $u,v$ leads to an equation for the function $F$ in the
form
\[
(\overline{F}^2-\theta\bar \theta)^2=\overline{F}(\overline{F}^2-3\theta\bar
\theta)+(s^2\bar \theta^3+
w^2\theta^3)\overline{F}
\]
where $s=\frac{\alpha}{M}$, $w=\frac{M^2r^2}{S}$ and
$\overline{F}\equiv F+4sw$.

\section{Conclusion}

The main result of the present paper is a parametric solution of the
Kepler problem in a deformed space whose construction involves
three parameters with dimensions of length, momentum and action
respectively.

The orbits of particles in this version of the Kepler problem turn
out to be (branches of) quartic curves. This should be compared
with the analysis of Higgs \cite{Higgs} who studies the orbits
in the case where there is only one non-trivial deformation parameter,
namely the length $L$. In Higgs' picture the orbits are elliptic,
parabolic or hyperbolic just as in the nondeformed case. This, however,
is due to a special choice of coordinates used in \cite{Higgs}.
These coordinates, though well-suited to the geometry of the problem,
have no physical significance. The coordinates we use are distinguished
among all possible coordinate systems as "the physical coordinates".
The transformation between Higgs' coordinates and the physical
coordinates is very simple: in our notations, Higgs' coordinates can
be written as $x_i/ I$.

The fact that the Kepler problem can be solved in the deformed space
gives some hope that other problems (for example, construction of
field theories) could also have reasonable solutions in deformed spaces.
The discussion of this subject is, however, beyond the scope of the
present article.

\appendix

\section*{Appendix: Structure of the algebra (\ref{commrel}).}
Throughout the section we assume that  $M^2>0$ and
\[\frac{1}{S^2}-\frac{1}{L^2M^2}>0.\]
Take the spherical coordinates with the angles
$(\phi,\theta,\chi)$ in $\bf{R}^4$.
Let $(\bs{u},u_4)$ be the unit vector
\[(\bs{u},u_4)=(\sin{\chi}\sin{\theta}\sin{\phi},\sin{\chi}\sin{\theta}\cos{\phi},
sin{\chi}\cos{\theta},\cos{\chi})\]
and set
\[p_{\chi}=\frac{\partial}{\partial\chi},
p_{\theta}=\frac{\partial}{\partial\theta}, p_{\phi}=\frac{\partial}{\partial\phi}.\]
Define the vectors $\bs{F}$ and $\bs{L}$ by
\[
\begin{array}{l}
F_1= \sin{\theta}\sin{\phi}\frac{\partial}{\partial\chi}+
\cot{\chi}\cos{\theta}\sin{\phi}\frac{\partial}{\partial\theta}+
\cot{\chi}\frac{\cos{\phi}}{\sin{\theta}}\frac{\partial}{\partial\phi},\\
F_2= \sin{\theta}\cos{\phi}\frac{\partial}{\partial\chi}+
\cot{\chi}\cos{\theta}\cos{\phi}\frac{\partial}{\partial\theta}-
\cot{\chi}\frac{\sin{\phi}}{\sin{\theta}}\frac{\partial}{\partial\phi},\\
F_3=\cos{\theta}\frac{\partial}{\partial\chi}+
\cot{\chi}\sin{\theta}\frac{\partial}{\partial\theta}
\end{array}
\]
and
\[
\begin{array}{l}
L_1=\sin{\phi}\frac{\partial}{\partial\theta}+\cos{\phi}\cot{\theta}\frac{\partial}{\partial\phi},\\
L_2=\cos{\phi}\frac{\partial}{\partial\theta}-\sin{\phi}\cot{\theta}\frac{\partial}{\partial\phi},\\
L_3=-\frac{\partial}{\partial\phi}.
\end{array}
\]
respectively.

Now,
\begin{align}
&I=\bs{(uF)}+ u_4\rho, \nonumber \\
&\frac{1}{M}(\bs{p}-\frac{M^2}{S}\bs{x})=\rho\bs{u}+\bs{[Lu]}-u_4\bs{F}, 
\nonumber\\[-.6em]
{} \\[-.6em]
&\bs{x}\cdot M\sqrt{\frac{1}{S^2}-\frac{1}{L^2M^2}}={\bs{F}},\nonumber \\
&\bs{l}\cdot\sqrt{\frac{1}{S^2}-\frac{1}{L^2M^2}}=\bs{L}. \nonumber
\end{align}

\section*{Acknowledgments}

We would like to thank Carlos Villegas-Blas for useful advice.
The second author was supported by the grant J-33616E awarded by CONACyT.

\bibliographystyle{amsplain}

\bigskip

\noindent A.Leznov: andrey@buzon.uaem.mx\\
J.Mostovoy: jacob@matcuer.unam.mx
\end{document}